\documentclass[twocolumn,showpacs,preprintnumbers,amsmath,amssymb,superscriptaddress]{revtex4}
\usepackage{graphicx}

\begin{document}

\title{Solitons in thermal media with periodic modulation of linear refractive index}


\author{Xuekai Ma}
\affiliation{Laboratory of Photonic Information Technology, South China Normal
University, Guangzhou 510631, P. R. China}
\author{Zhenjun Yang}
\affiliation{Laboratory of Photonic Information Technology, South China Normal
University, Guangzhou 510631, P. R. China}
\author{Daquan Lu}
\affiliation{Laboratory of Photonic Information Technology, South China Normal
University, Guangzhou 510631, P. R. China}
\author{Wei Hu}
\affiliation{Laboratory of Photonic Information Technology, South China Normal
University, Guangzhou 510631, P. R. China}
\affiliation{State Key Laboratory of
Transient Optics and Photonics, Chinese Academy of Sciences, Xi'an 710119, P.
R. China} \affiliation{Corresponding author: huwei@scnu.edu.cn}

\begin{abstract}We address the existence and properties of solitons in
thermal media with periodic modulation of linear refractive index.
Many kinds of solitons in such optical lattices, including symmetric
and antisymmetric lattices, are found under different conditions. We
study the influence of the refractive index difference between two
different layers on solitons. It is also found that there do not
exist cutoff value of propagation constant and soliton power for
shifted lattice solitons. In addition, the solitons launched away
from their stationary position may propagate without oscillation
when the confinement from lattices is strong.
\end{abstract}



\maketitle 

\section{Introduction}
Nonlocal solitons have been studied in many physical systems, such
as photorefractive crystals~\cite{Mitchell-1998-PRL}, nematic liquid
crystals~\cite{Conti-2003-PRL,Conti-2004-PRL,Peccianti-2002-OL,Krolikowski-2005-PRe,Rasmussen-2005-PRE},
lead
glasses~\cite{Rotschild-2005-PRL,Rotschild-2006-NP,Skupin-2006-PRE},
Kerr-like media~\cite{Krolikowski-2004-job}, materials with
quadratic
nonlinearities~\cite{Nikolov-2003-PRE,Larsen-2006-PRE,Bache-2007-OL,Bache-2008-OE},
atomic vapors~\cite{Skupin-2007-PRL}, and Bose-Einstein
condensates~\cite{Pedri-2005-PRL,Tikhonenkov-2008-PRL}, etc. They
show some special and novel properties both in experiment and
theory, for instant, the large phase shift~\cite{Guo-2004-PRE},
self-induced fractional Fourier transform~\cite{Lu-2008-PRA},
attraction between two dark
solitons~\cite{Nikolov-2004-OL,Dreischuh-2006-PRL}, etc. Recently,
nonlocal surface solitons propagating along  the interface formed by
a nonlocal nonlinear medium and a linear medium are found and
observed~\cite{Alfassi-2007-PRL,Ye-2008-PRA,Kartashov-2009-OL,Alfassi-2009-PRA}.
This kind of solitons presents strong asymmetry including their
profiles and nonlinear refractive indices because of the different
boundary conditions. They are stable only when the number of
intensity peaks is less than three~\cite{Kartashov-2009-OL}.

Solitons are also found in some media with periodic
nonlinearity~\cite{Sukhorukov-2000-PRE,Sakaguchi-2005-PRE,abdullaev-2005-PRAR,bludov-2006-PRA,beitia-2007-PRL}.
Recently, solitons in nonlocal nonlinear lattices are studied, such
as odd, even, dipole, tripole and twisted
solitons~\cite{xu-2005-PRL,Kartashov-2008-OL1,Efremidis-2008-PRA,Kartashov-2011-RMP}.
For lattice solitons formed in nonlocal media with an imprinted
modulation of linear refractive index, there exists a cutoff point
of propagation constant due to the existence of the band-gap
structure of periodic lattice~\cite{xu-2005-PRL}. In thermal
lattices with periodic nonlinearity, there can be found asymmetric
solitons, which is shifted solitons. There exist a cutoff value of
propagation constant and a minimal power for this kind of solitons
in such lattices~\cite{Kartashov-2008-OL1}. It is found that, in
thermal nonlinear lattices, solitons can exist anywhere inside a
band and two families of solutions bifurcate from each band edge
because of the infinite range of nonlocality and the presence of
boundaries~\cite{Efremidis-2008-PRA}. At the same time, surface
lattice solitons are found at the interface between a uniform medium
and a nonlinear lattice or at the edge of layered nonlocal
media~\cite{Kartashov-2006-OL,Kartashov-2009-OL}. Such solitons
present strongly asymmetric shapes and they are stable in large
parts of their existence region. There is no restriction on the
number of peaks in stable surface solitons localized at a thermally
insulating interface between layered thermal media and a linear
dielectric, which is quite different from that in uniform nonlocal
media~\cite{Kartashov-2009-OL}.

Interface solitons propagating at the interface of one-dimensional
thermal nonlinear media with a step in linear refractive index have
been studied in \cite{Ma-2011-PRA}. It is note that interface
solitons in such media are asymmetric because of the difference
of the linear refractive index. For tripole and quadrupole interface
solitons, there exist two different types of solutions under some
special conditions~\cite{Ma-2011-toPRA}.

In this paper, we study the propagation of solitons in the
semi-infinite band gap of thermal media with periodic distribution
of linear refractive index. In our model, the period modulation
function of linear index is a step function. We find the odd, even,
and shifted solitons in this kind of media. In addition, there do
not exist cutoff value of propagation constant for shifted solitons,
which is different from that in thermal media with periodic
nonlinearity. We also study the influence of the index-difference of
the lattices and the boundaries on solitons.

\section{Theoretical Model}
We consider a (1+1)-dimensional layered thermal sample occupying the
region $-L\leq x\leq L$ imprinted a linear refractive index lattice.
A laser beam propagating along the $z$ axis is governed by the
nonlocal nonlinear Schr\"{o}dinger equation (NNLSE) for the
dimensionless amplitude $q$ of the light field, coupled to the
equation for normalized nonlinear refractive index variation $n$,


\begin{equation}\label{1}
  i\frac{\partial q}{\partial
z}+\frac{1}{2}\frac{\partial^2q}{\partial x^2}+nq+\delta(x)n_dq=0,
\end{equation}
\begin{equation}\label{11}
  \frac{\partial^2n}{\partial x^2}=-|q|^2,
\end{equation}
where $x$ and $z$ stand for the transverse and longitudinal
coordinates scaled to the beam width and the diffraction length,
$n_d$ represents the difference between the linear refractive
indices, $\delta(x)$ is a periodic function for the modulation of
the linear refractive index. In our model, we study the cases of
$\delta(x)=H[\cos(\pi x/d)]$ and $\delta(x)=H[\sin(\pi x/d)]$, which
present the symmetric and antisymmetric lattices, respectively. Here
$H[\xi]$ is the Heaviside step function defined as
\begin{equation}\label{2}
H[\xi]=\begin{cases}
1, & \xi \geq 0,\\
0, & \xi <0.
\end{cases}
\end{equation}
$d$ is the layer width and the lattice period is $2d$. For
simplicity and without losing of generality, we use integer value of
$d$ in this paper, which possibly leads to the incomplete layers at
the two boundaries. Anyway, the forth term of Eq. (\ref{1})
represents the periodic modulation of two different linear
refractive index in the sample. This kind of modulation is different
from that in Ref.~\cite{Efremidis-2008-PRA}. When $\delta(x)=1$, it
represents the higher linear refractive index, while $\delta(x)=0$,
it represents the lower linear refractive index. If $n_d=0$, Eq.
(\ref{1}) represents the NNLSE in uniform thermal media.

We assume the two boundaries are thermally conductive and the laser
beam is far away from the boundaries, so the boundary conditions are
$q(\pm L)=0$ and $n(\pm L)=0$. We search for soliton solutions of
Eq. (\ref{1}) numerically in the form $q(x,z)=w(x)\exp(ibz)$, where
$w(x)$ is a real function, $b$ is the propagation constant. The
iterative method is used to get numerical solutions for different
$n_d$, $d$ and $b$ as shown in Figs. \ref{f1}-\ref{shifted}. To
elucidate the linear stability, we search for the perturbed solution
of Eq. (\ref{1}) in the form $q=(w+u+iv)\exp(ibz)$, the real part
$u(x,z)$ and the imaginary part $v(x,z)$ of the perturbation can
grow with a complex rate $\sigma$ upon propagation. Substituting the
perturbed solution into Eq. (\ref{1}), one can get the linear
eigenvalue problem around stationary solution $w(x)$,
\begin{equation}\label{3}
 \left.\begin{aligned}
  \sigma u&=-\frac{1}{2}\frac{d^2v}{dx^2}+bv-nv-\delta(x)n_dv,  \\
  \sigma v&=\frac{1}{2}\frac{d^2u}{dx^2}-bu+nu+\delta(x)n_du+w\Delta n,  \\
\end{aligned}\right\}
\end{equation}
where $\Delta n=-2\int_{-L}^{L}G(x,x')w(x')u(x')dx'$ is the
refractive index perturbation, the response function
$G(x,x')=(x+L)(x'-L)/(2L)$ for $x\leq x'$ and
$G(x,x')=(x'+L)(x-L)/(2L)$ for $x\geq x'$. $\sigma_r$ (real part of
$\sigma$) represents the instability growth rate.

\section{Lattice Solitons}

\begin{figure}[t]
\centerline{\includegraphics[width=8.0cm]{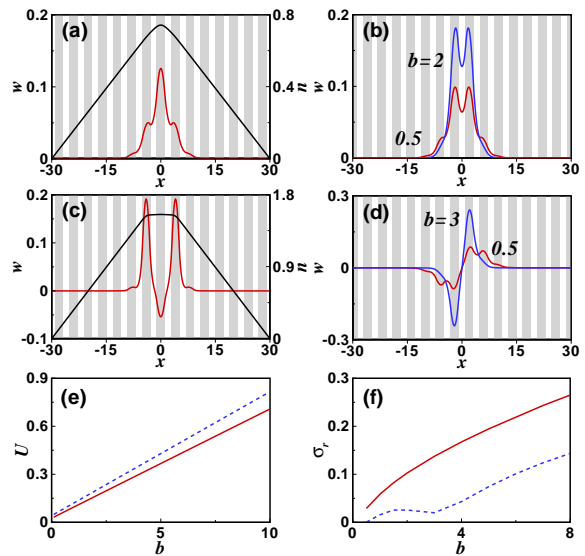}}
 \caption{(Color online) Profiles of a
fundamental soliton (a) at $n_d=1$, $b=0.5$, even solitons (b) at $n_d=-0.5$, a
tripole soliton (c) at $n_d=2$, $b=1$, and dipole solitons (d) at $n_d=-0.5$.
(e) Energy flows versus propagation constant of (a) (solid red line) and (c)
(dashed blue line). (f) Real part of the perturbation growth rate of (b) (solid
red line) and (d) (dashed blue line). In all cases $d=2$.}\label{f1}
\end{figure}

First, we study the solitons in the case of $\delta(x)=H[\cos(\pi
x/d)]$ (symmetric lattices) and $d=2$ as shown in Fig. \ref{f1}.
When $n_d>0$ the center layer is with higher refractive index, and
the number of higher-index layers is odd. In Figs. \ref{f1}(a) and
\ref{f1}(c), there exist odd solitons in the center of the sample
because intensity peaks always reside in the higher-index of the
layers. Their profiles are similar to those in other types of
optical lattices, while the distribution of nonlinear refractive
indices are different~\cite{xu-2005-PRL,Kartashov-2008-OL1}. When
$n_d<0$ the center layer is with lower  refractive index, and the
number of higher-index layers is even. In Figs. \ref{f1}(b) and
\ref{f1}(d), one can find even solitons, including symmetric ones
[Fig. \ref{f1}(b)] and antisymmetric ones [Fig. \ref{f1}(d)], at the
center of the sample. As we know that the beam width decreases as
the propagation constant increases. In optical lattices, the smaller
the $b$ is, the more layers the soliton profile occupies. That is
the number of the intensity peaks decreases as the propagation
constant increases [Figs. \ref{1}(b) and \ref{1}(d)].

The linear stability analysis shows that fundamental and tripole
lattice solitons in Figs. \ref{f1}(a) and \ref{f1}(c) are stable in
their existence domain for $n_d>0$. Energy flows
$U=\int_{-\infty}^{\infty}|q|^2dx$ of fundamental and tripole
solitons are monotonically growing functions of $b$ as shown in Fig.
\ref{f1}(e). However, when $n_d<0$, the even and dipole solitons, as
shown in Figs. \ref{1}(b) and \ref{1}(d), are unstable for
$n_d=-0.5$ [Fig. \ref{f1}(f)]. We find that the antisymmetric dipole
solitons as shown in Fig.~\ref{1}(d) can be stable for $n_d< -0.6$
due to the confinement of the deep lattice, whereas the even
solitons as shown in Fig. \ref{1}(b) are unstable for any value of
$n_d$.

\begin{figure}[t]
\centerline{\includegraphics[width=8.0cm]{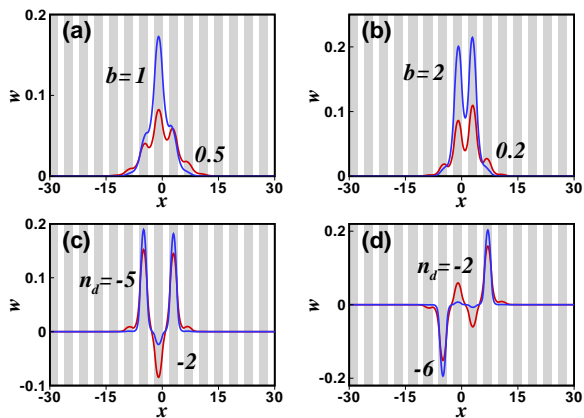}}
\caption{(Color online) Solitons in antisymmetric lattices at (a)
$n_d=-0.5$, (b) $n_d=-1$, (c) and (d) $b=0.5$. In all cases $d=2$
and $L=30$.}\label{odd}
\end{figure}

Then, we discuss the solitons formed in the antisymmetric lattices.
The modulation of linear refractive index is governed by
$\delta(x)=H[\sin(\pi x/d)]$ and lattice structure is shown in Fig.
\ref{odd} with $d=2$. If there is only one period in the sample,
i.e. $d=L$, the antisymmetric lattice will be the same as the model
in \cite{Ma-2011-PRA}. Here solitons are always asymmetric because
of the antisymmetry of lattices. Figures \ref{odd}(a) and
\ref{odd}(b) show the fundamental and even asymmetric solitons,
respectively. One can see that the main intensity peaks reside in
the higher-index layers, and the larger the propagation constant the
higher the degree of symmetry. Figures \ref{odd}(c) and \ref{odd}(d)
show the high-order asymmetric solitons at the same propagation
constant. One can see that the degree of the restriction of the
lattices is proportional to $|n_d|$, and the asymmetry of the
solitons is almost independent of $n_d$. The fundamental and the
high-order asymmetric solitons shown in Figs. \ref{odd}(a),
\ref{odd}(c) and \ref{odd}(d) are stable in their whole existence
domain, while the even solitons shown in Fig. \ref{odd}(b) are
unstable.

\begin{figure}[t]
\centerline{\includegraphics[width=8.0cm]{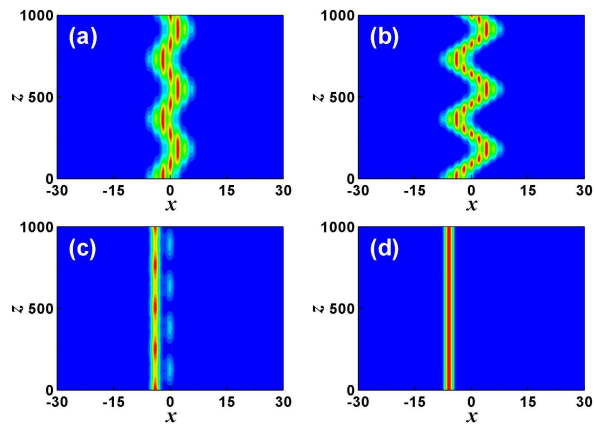}}
\caption{(Color online) Propagations of solitons launched away from
the sample center for (a) and (b) $b=0.5$, $n_d=1$, $d=1$, (c)
$b=1$, $n_d=1$, $d=2$, and (d) $b=0.5$, $n_d=3$, $d=3$. In all cases
$L=30$.}\label{excenter}
\end{figure}

As we know in~\cite{Alfassi-2007-OL}, a soliton launched away from
the sample center will oscillate periodically because of the
restoring boundary force. In our model, we simulate the beam
propagations with finite difference method in the symmetric
lattices. Solitons launched away from the sample center undergo
oscillation, as shown in Figs. \ref{excenter}(a) and
\ref{excenter}(b), when the boundary force is stronger than the
constraint force of the higher-index waveguide~\cite{dai-2008-PRA}.
It is important to note that the intensity peaks jump between the
adjoining higher-index layers during propagations because the peaks
will not reside in the lower-index layers, which is the same as
discrete diffraction \cite{diffraction}. If the soliton is launched
further away from the sample center as shown in Fig.
\ref{excenter}(b), it also oscillates periodically with larger span,
but its oscillating period is the same as that in Fig.
\ref{excenter}(a).

If we increase the layer width [Fig. \ref{excenter}(c)], or the
difference between the two linear refractive indices [Fig.
\ref{excenter}(d)], the solitons will propagate without oscillation.
Similar to the results in \cite{dai-2008-PRA}, when the constraint
force is stronger than the restoring force, the intensity peak is
difficult to jump into other higher-index layers for large $n_d$ and
$d$. It is worth to note that only fundamental solitons will
oscillate periodically, the high-order solitons will oscillate
irregularly and destroy themselves for small $n_d$ and $d$. However,
stationary high-order solitons can form for large $n_d$ and $d$ when
they are launched away from the sample center.

\begin{figure}[b]
\centerline{\includegraphics[width=8.0cm]{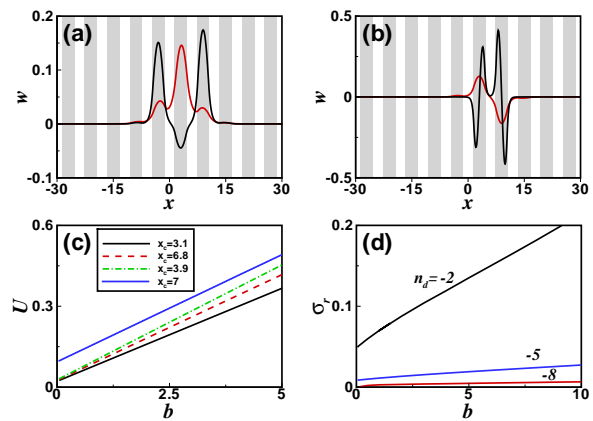}}
\caption{(Color online) Profiles of (a) fundamental and tripole with
$b=0.05$, $n_d=2$, and $d=3$; (b) dipole and quadrupole shifted
solitons with $b=0.05$, $n_d=-2$, and $d=3$. (c) Energy flows versus
propagation constant of fundamental, dipole, tripole, and quadrupole
solitons. (d) Real part of the perturbation growth rate of the
quadrupole soliton.}\label{shifted}
\end{figure}

\begin{figure*}[htbp]
\centerline{\includegraphics[width=12cm]{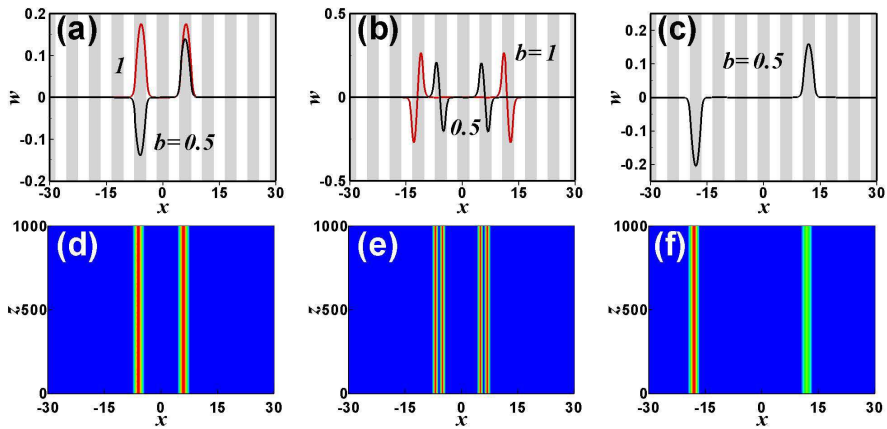}}
 \caption{(Color online) Profiles of
shifted solitons at (a) $b=0.5$ (black line) and $b=1$ (red line),
(b) $b=0.5$ (black line) and $b=1$ (red line), and (c) $b=0.5$. In
all cases $n_d=5$. (d)-(f) Propagations of shifted solitons
corresponding to (a)-(c) at $b=0.5$.}\label{shifted2}
\end{figure*}

Based on the above analysis, when the constraint force of the
lattices is large, solitons can form asymmetrically away from the
sample center, that is shifted solitons. Figures \ref{shifted}(a)
and \ref{shifted}(b) show the profiles of fundamental, dipole,
tripole, and quadrupole shifted solitons at the same conditions.
Their mass center, defined as
$x_m=\int_{-\infty}^{\infty}x|q|^2dx/\int_{-\infty}^{\infty}|q|^2dx$,
are $3.1$, $6.8$, $3.9$, and $7$, respectively. This kind of
solitons exists when the constraint  force of the higher-index
waveguide is stronger than the restoring force exerted by two
boundaries due to the shift of the soliton center. Figure
\ref{shifted}(c) shows that their energy flows are monotonically
growing function of $b$ and there do not exist a cutoff value of $b$
for any shifted solitons. It is found that there exist a minimal
power and a cutoff value of $b$ for shifted solitons in the model in
\cite{Kartashov-2008-OL1}, because the modulation in their model is
induced by the nonlinearity, which will become weak for small
propagation constant and low soliton power. However, in our model,
the modulation of linear refractive index does not depend on the
constant propagation, so there does not exist a cutoff value of $b$
for shifted solitons. For fundamental, dipole, and tripole shifted
solitons shown in Figs. \ref{shifted}(a) and \ref{shifted}(b), they
are all stable in their whole existence domain, while for the
quadrupole shifted soliton, there exists an instability region shown
in Fig. \ref{shifted}(d). We can see that the instability region is
dependent on $n_d$, and the instability region decreases as
increasing of $|n_d|$. As $|n_d|>8$, almost all solitons are stable.

Another type of shifted solitons formed away from the sample center
is shown in Fig. \ref{shifted2}. They are separated  by the central
layers and located in any higher-index layer, including symmetrical
[the red lines in Figs. \ref{shifted2}(a) and \ref{shifted2}(b)],
antisymmetrical [the black lines in Figs. \ref{shifted2}(a)and
\ref{shifted2}(b)], and asymmetrical [Fig. \ref{shifted2}(c)] ones,
because of the restraint force of lattices and the existence of
boundaries. Almost all intensity peaks reside in the higher-index
layers for this type of solitons, and they do not affect other peaks
during the propagation due to the restriction of the lattices. For
this reason, the solitons shown in Figs.
\ref{shifted2}(a)-\ref{shifted2}(c) are stable including the
quadrupole shifted solitons, which is different form that in Fig.
\ref{shifted}(b). Their propagations are shown in Figs.
\ref{shifted2}(d)-\ref{shifted2}(f). One can see from Fig.
\ref{shifted2}(f) that the two intensity peaks are stable and have
no impact on each other, even if they are far away from the sample
center.

\begin{figure}[b]
\centerline{\includegraphics[width=8.0cm]{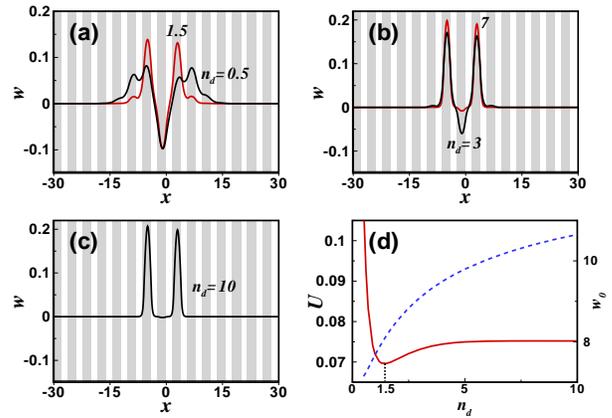}} \caption{(Color
online) Profiles of solitons at (a) $n_d=0.5$ and $n_d=1.5$, (b)
$n_d=3$ and $n_d=7$, and (c) $n_d=10$. In all cases $d=2$ and
$b=0.5$. (d) Energy flow (blue dashed line) and beam width (red
solid line) versus $n_d$.}\label{nd}
\end{figure}

Finally, we discuss the influence of the index-difference on
solitons with $d$ and $b$ fixed in the antisymmetrical lattices.
When $n_d$ is small, as shown in Fig. \ref{nd}(a), there are three
main intensity peaks with a significant part of the field residing
in the lower-index layers and exist several sub-peaks. That is
because the higher-index layers present a little restriction at
small $n_d$. As $n_d$ increases [Fig. \ref{nd}(b)], the part of the
field residing in the lower-index layers decreases, and the middle
intensity peak decreases gradually. When $n_d=10$, almost all
soliton power resides in the higher-index layer and the middle peak
disappears as shown in Fig. \ref{nd}(c). It shows that the
restriction of the higher-index layers increases as $n_d$ increases.
Figure \ref{nd}(d) shows the relationship between $n_d$ and beam
width and the energy flow. One can see that the energy flow is
proportional to $n_d$, while the beam width changes
non-monotonically with increasing $n_d$. When $n_d\leq1.5$, the beam
width decreases monotonically as $n_d$ increases as shown in Fig.
\ref{shifted2}(d). The main reason is that the widthes of two
lateral peaks decrease as $n_d$ increases, while the width of the
middle peak is unchanged [Fig. \ref{shifted2}(a)]. When
$n_d\geq1.5$, the beam width increases monotonically as $n_d$
increases [Fig. \ref{shifted2}(d)] because the width of the middle
peak decreases gradually, while the widthes of two lateral peaks are
unchanged with increasing $n_d$. When $n_d>5$, the beam width
approaches $5$ because the middle peak almost disappears and there
is little change of the soliton profile any more [Fig.
\ref{shifted2}(b) and Fig. \ref{shifted2}(c)]. For comparison,
soliton profiles in symmetric lattices are symmetric when the
constraint force is weaker than the restoring force [Figs.
\ref{f1}], whereas they can be either symmetric [Fig.
\ref{shifted2}] or asymmetric [Figs. \ref{shifted} and
\ref{shifted2}] when the constraint force is stronger than the
restoring boundary force for large $n_d$ [Fig. \ref{shifted}]. For
solitons in antisymmetric lattices, however, they are always
asymmetric [Figs. \ref{odd} and \ref{nd}].

\section{Conclusion}
To conclude, we have studied the propagation of solitons in thermal
media with periodic distribution of linear refractive index
including symmetric and antisymmetric lattices. In this kind of
optical lattice, we found fundamental, odd, even, and shifted
solitons. The shifted solitons are found do not exist cutoff value
of propagation constant and power, which is different from that in
thermal lattices with periodic nonlinearity. We also studied the
solitons launched away from their stationary position. Their
propagation is related to the difference of the refractive indices.

\section*{Acknowledge}
This research was supported by the National Natural Science
Foundation of China (Grant Nos. 10804033, 10674050, 11174090, and
11174091), the Specialized Research Fund for the Doctoral Program of
Higher Education (Grant No.200805740002), and the Open Research Fund
of State Key Laboratory of Transient Optics and Photonics, Chinese
Academy of Sciences.

\end{document}